\documentclass[pra,twocolumn,showpacs,preprintnumbers,amsmath,amssymb,superscriptaddress]{revtex4}
\usepackage{graphicx}% Include figure files
\usepackage{bm}% bold math
\usepackage{commath}
\usepackage{color}
\usepackage{amsmath}

\begin{document}

%\preprint{APS/123-QED}

\title{Covariance matrix entanglement criterion for an arbitrary set of operators}

\author{Vinay Tripathi}
\affiliation{New York University Shanghai, 1555 Century Ave, Pudong, Shanghai 200122, China}

\author{Chandrashekar Radhakrishnan}
\affiliation{New York University Shanghai, 1555 Century Ave, Pudong, Shanghai 200122, China}

\author{Tim Byrnes\footnote{tim.byrnes@nyu.edu}}
%\affiliation{State Key Laboratory of Precision Spectroscopy, School of Physical and Material Sciences,East China Normal University, Shanghai 200062, China}
%\affiliation{NYU-ECNU Institute of Physics at NYU Shanghai, 3663 Zhongshan Road North, Shanghai 200062, China}
%\affiliation{National Institute of Informatics, 2-1-2 Hitotsubashi, Chiyoda-ku, Tokyo 101-8430, Japan}
\affiliation{New York University Shanghai, 1555 Century Ave, Pudong, Shanghai 200122, China}
\affiliation{Department of Physics, New York University, New York, NY 10003, USA}

\date{\today}% It is always \today, today,
\begin{abstract}
We generalize entanglement detection with covariance matrices for an arbitrary set of observables.
A generalized uncertainty relation is constructed using the covariance and commutation
matrices, then a criterion is established by performing a partial transposition on
the operators. The method is highly efficient and versatile in the sense that the set of measurement
operators can be freely chosen, do not need to be complete, and there is no constraint
on the commutation relations. The method is particularly suited for systems with higher
dimensionality since the computations do not scale with the dimension of the Hilbert space
— rather they scale with the number of chosen observables which can always be kept small.
We illustrate the approach by examining the entanglement between two spin ensembles, and
show that it detects entanglement in a basis independent way. 
\end{abstract}

\pacs{03.75.Dg, 37.25.+k, 03.75.Mn}% PACS, the Physics and Astronomy
\maketitle

\section{Introduction}
Entanglement is a key resource for many quantum information technologies and detecting its presence is a fundamental experimental task.  Numerous methods for the detection and quantification of entanglement have been studied in great depth for bipartite and multipartite systems \cite{horod,guhne2009entanglement,eisert2007quantitative,vedral2008quantifying,plenio2005introduction}.  For systems with small Hilbert space dimension, one standard approach is to reconstruct the density matrix and perform a positive partial transpose (PPT) test to check for separability \cite{peres,Horodecki19961,HORODECKI1997333,PhysRevLett.95.090503,vidal2002computable}.  However, this requires full tomography of the density matrix, which for systems with large Hilbert space may be impractical or even impossible to measure.  In this case what is most desirable (e.g. from an experimental point of view) are simple criteria that can be evaluated based on a small number of observables.  In this sense, criteria such as those given by Duan and co-workers \cite{Duan}, Hillery-Zubairy \cite{hillery}, entanglement witness \cite{horedecki1996separability,PhysRevA.62.052310,szangolies2015detecting,singh2016entanglement}, mutually unbiased bases \cite{PhysRevA.86.022311}, and others \cite{toth2006detection,PhysRevA.92.022339,wu2005entanglement,PhysRevA.95.052305,de2011multipartite,altepeter2005experimental} are quite valuable as they can detect entanglement without full tomography of the density matrix.

For continuous variables, a powerful method to detect entanglement has been developed based on the covariance matrix  \cite{simon,Duan,braunstein,adesso2007entanglement,wang2007quantum,werner}.  In the approach, one has observables corresponding to the quadrature pairs in two subsystems (labeled by $A$ and $B$) $ \xi = (x_A, p_A, x_B, p_B) $ where $ [x,p] = i  $. Defining the covariance matrix, one may determine a sufficient condition for entanglement \cite{simon,Duan}.  Despite its great success for optical systems, for other systems, it has been more challenging to define analogous quantities. Generalization to nonlinear systems \cite{guhne2007nonlinear} and necessary and sufficient inseparability conditions for symmetric qubits have been performed \citep{usha,gittsov}. In particular, a more general framework for finite systems and a generalized set of measurement operators was performed by G{\"u}hne, Eisert, and co-workers, with several entanglement criteria defined accordingly \cite{guhne2007,PhysRevA.81.032333}.  While this allows for a more generalized application of the approach, the approach requires a complete set of observables to construct the covariance matrix \cite{li2008separability}.  For a large but finite systems, such as two atomic spin ensembles or Bose-Einstein condensates, this makes it difficult to apply in practice, since measurements scaling as the square of the dimension of the Hilbert space are needed \cite{polzik,andrew,eisert}. What would be more desirable is to have a covariance matrix approach that is based on a finite set of freely choosable measurements, and applicable to any system, both finite and infinite. 

In this paper, we generalize the covariance matrix approach to an arbitrary set of measurement operators, that satisfy arbitrary commutation relations. The situation that is most relevant to our formalism is as follows.  Consider that a set of correlations $ \langle \xi_j \xi_k \rangle $  and expectation values $ \langle \xi_j \rangle $ has been measured (e.g. from an experiment), where $ \xi_j $ are an arbitrary set of $ N $ known observables.  The task is then to take this data and determine whether entanglement is present between two subsystems, $A $ and $B$.  
In the situation we consider, we assume that an arbitrary entanglement witness operator is not in general constructable  using the $ \xi_j $ operators, hence the question is how best to extract the information regarding entanglement from the given correlations.   Our approach is to construct a covariance matrix and a commutation matrix, which together can be used to determine the presence of entanglement.  Since the number of observables $ N $ is typically quite small, this is a highly efficient procedure for detecting entanglement since it only requires diagonalization of a $ N \times N $ matrix.  This is in contrast to alternative entanglement detection methods which can require computations scaling with the Hilbert space dimension $ D \gg N $, which can potentially be large.

\section{Covariance and commutator matrix}
We start with defining a set of $ N $ observables (Hermitian operators) $ \xi_n $ on the composite system $ A $ and $ B $, with $ n \in [1, N] $. The dimension of the Hilbert space of the composite system is $ D $.   All these operators can be collected in a vector
\begin{align}
\xi = ({\cal \xi}_1,{\cal \xi}_2 \dots, {\cal \xi}_{N})
\label{xivector}
\end{align}
where these operators are general so that they could come from either subsystem $ A $, $ B $, or both.  
The $ N \times N $  covariance matrix is defined in the standard way  
\begin{align}
V_{jk} \equiv  \frac{1}{2} \langle \{ \xi_j, \xi_k \} \rangle - \langle \xi_j \rangle  \langle \xi_k \rangle ,  
\end{align}
which is a real symmetric matrix and $ \{ X, Y \} = XY + YX $ is the anticommutator. The $N \times N$  commutation matrix for  $ j,k \in [1,N] $ is defined as
\begin{align}
\Omega_{jk} \equiv  -i\langle [\xi_j,\xi_k] \rangle,
\label{omega}
\end{align}
which is a real antisymmetric matrix \footnote{See Appendix.}.  

The matrix inequation 
\begin{align}
 V+\frac{i}{2}\Omega \ge 0 
\label{Simon1}
\end{align}
succinctly summarizes the uncertainty relation between the operators $ \xi_j$ \cite{simon,simonearlypaper}.  For the purposes of the relation (\ref{Simon1}) there is no role played by the subsystems $ A, B $ and hence we may consider $ \xi $ to contain an arbitrary set of $ N $ operators. 
The meaning of (\ref{Simon1}) is in terms of the semi-positive nature of the matrix, i.e. that it has no negative eigenvalues.  In Ref. \cite{simonearlypaper} this was shown for the case of quadratures.  We show here that in fact this generalizes to arbitrary number and type of operators.  To see the power of the relation (\ref{Simon1}) consider the generalized uncertainty relation for an arbitrary number of operators. The Schrodinger uncertainty relation gives the relationship bounding the product of the variances between two operators $ \xi_1, \xi_2 $
\begin{align}
{\cal I}_{12} \equiv & \sigma^2_{\xi_1}\sigma^2_{\xi_2} - \left| \frac{\langle\{\xi_1,\xi_2\}\rangle}{2} -\langle \xi_1 \rangle \langle \xi_2 \rangle \right|^2   - \left|  \frac{\langle[\xi_1 ,\xi_2]\rangle}{2i} \right|^2 \ge 0  , \nonumber
\end{align} 
where $ \sigma^2_\xi \equiv \langle \xi^2 \rangle - \langle \xi \rangle^2 $. Following the same procedure \footnotemark[53], one can straightforwardly derive the Schrodinger uncertainty relations for $ N $ operators  $ \xi_1, \dots, \xi_N $.  For example, for $ N =1 $ one obtains $ {\cal I}_{1} \equiv \sigma^2_{\xi_1} \ge 0 $ and for $ N = 3 $ we obtain
\begin{align}
{\cal I}_{123} \equiv & \sigma^2_{\xi_1}\sigma^2_{\xi_2}\sigma^2_{\xi_3} - \langle{f_1}|f_1\rangle|\langle{f_2}|f_3\rangle|^2 -\langle{f_2}|f_2\rangle|\langle{f_3}|f_1\rangle|^2  \nonumber \\ 
&  - \langle{f_1}|f_2\rangle\langle{f_2}|f_3\rangle\langle{f_3}|f_1\rangle   -\langle{f_2}|f_1\rangle\langle{f_3}|f_2\rangle\langle{f_1}|f_3\rangle   &\nonumber \\
&- \langle{f_3}|f_3\rangle|\langle{f_1}|f_2\rangle|^2  \ge 0 
  \label{ee}
\end{align} 
where  $ |f_i\rangle=(\xi_i -\langle \xi_i \rangle)|\Psi\rangle $. 

   The remarkable feature of (\ref{Simon1}) is that it contains information about all the Schrodinger uncertainty 
relations as described above. Taking for example the $ N = 3 $ case, the first order invariant of (\ref{Simon1}) yields $  {\cal I}_{1} +  {\cal I}_{2}  +  {\cal I}_{3} \ge 0 $, i.e. the sum of the variances of the operators is non-negative.  The second order invariant (sum of principle minors) yields $  {\cal I}_{12} +  {\cal I}_{23}  +  {\cal I}_{13} \ge 0 $, which is the sum of the standard Schrodinger uncertainty relation between all operator pairs.  Finally, the third order invariant (i.e. the determinant) yields $ {\cal I}_{123} \ge 0  $, which is the three operator Schrodinger uncertainty relation.  In the $ N $-operator case,  (\ref{Simon1}) summarizes the $ 1, 2, \dots, N $-operator Schrodinger uncertainty relation via the matrix invariants in a highly succinct way.  

While this connection to the uncertainty relation is beautiful, we should show explicitly that (\ref{Simon1}) is true for an arbitrary set of operators.  In fact there is a simple way to show this in general without alluding to the uncertainty relation.  Let us first write 
\begin{equation}
V +\frac{i}{2}\Omega =  
\begin{pmatrix}
\langle{f_1}|f_1\rangle & \ldots & \langle{f_1}|f_N\rangle\\
\vdots & \ddots & \vdots\\
\langle{f_N}|f_1\rangle & \ldots & \langle{f_N}|f_N\rangle\\
\end{pmatrix} .
\label{V2}
\end{equation}
This an overlap matrix, which is positive definite, immediately implying that
$ V+\frac{i}{2}\Omega > 0 $ for pure states \cite{overlap}. Overlap matrices typically assume that none of the matrix elements are zero. Here we relax the condition and allow for the possibility that $ \langle{f_j}|f_k\rangle = 0 $.  This gives the additional possibility that the matrix can possess a zero eigenvalue, which shows the positive semi-definite nature of (\ref{Simon1}).  For mixed states, we may use the fact that any density matrix can be diagonalized into a mixture of pure states $ \rho = \sum_l p_l | \Psi_l \rangle \langle  \Psi_l |$ and the concavity property of covariance matrices \cite{gitts} to show that (\ref{Simon1}) is true for mixed states \footnotemark[53].

The matrix formalism is convenient as it automatically takes into account of symmetries that may be present in the operators $ \xi_j $.  The uncertainty relations are in terms of invariants of the matrix, hence they are guaranteed to be unchanged under these symmetry operations.   For example, for the quantum optical case with operators $\xi = (x_A,p_A,x_B,p_B) $, symplectic transformations Sp(4) can be used to obtain another set of observables $\xi' = (x_A',p_A',x_B',p_B') $ \citep{simonearlypaper}. The uncertainty relations are guaranteed to be unchanged under such a transformation.  For the case of a spin ensemble that we examine later, we use observables $\xi = (S_x,S_y,S_z) $, which can be rotated under a basis transformation under SO(3) to $\xi' = (S_x',S_y',S_z') $.  Using a symmetrical set of operators, the particular basis that is used for the measurements becomes irrelevant.

\section{Entanglement detection}
Now that we have established (\ref{Simon1}), we may see how this can be used in relation to detecting entanglement. 
%Let us now assume a bipartite structure to (\ref{xivector}), such that the first $ N_A $ operators are in subsystem $ A $ and the remaining $ N_B $ are in subsystem $ B $:
%
%\begin{align}
%\xi = (\xi_1^{(A)}, \dots, \xi_{N_A}^{(A)}, \xi_{N_A+1}^{(B)}, \dots,  \xi_{N}^{(B)} )
%\label{bipartitexi}
%\end{align}
%
%where we have $ N = N_A + N_B $. 
We use the Peres-Horodecki criterion \citep{peres,HORODECKI1997333,Horodecki19961} by performing a partial transpose (PT) operation on the density matrix operator. This PT operation on the density  matrix operator will have a corresponding effect on the covariance matrix $ V \rightarrow \text{PT} (V) $, which depends upon  the particular operators that are used in $ \xi $. The commutation matrix also is affected in the general case according to 
$ \Omega \rightarrow \text{PT} (\Omega) $.  Denoting the PT transposed density matrix as $ \rho^{T_B} $, The PT transposition operation is defined as making the replacement $ \rho \rightarrow \rho^{T_B} $ in all averages:
\begin{align}
[\text{PT} (V)]_{jk} &= \frac{1}{2}\text{Tr} (\rho^{T_B}\{\xi_j,\xi_k \}) - \text{Tr} (\rho^{T_B}\xi_j )  \text{Tr} (\rho^{T_B} \xi_k )  \nonumber  \\
[\text{PT} (\Omega)]_{jk} &= -i\text{Tr} (\rho^{T_B} [ \xi_j, \xi_k])  .
\label{pttransformed}
\end{align}
For a separable state, the PT operation should give a valid density matrix with positive eigenvalues.  Thus the new covariance and commutation matrix necessarily satisfies
\begin{align}
\text{PT} (V) + \frac{i}{2} \text{PT} (\Omega)\ge 0 
\label{necessary}
\end{align}  
if it is a separable bipartite state.  Violation of (\ref{necessary}) guarantees non-separability and thus entanglement in the bipartite system. 

The usefulness of the above argument hinges upon the simple evaluation of (\ref{pttransformed}).  The partial transposed operator is not available by direct measurement, hence we require an equivalent expression in terms of the original density matrix. Due to the fact that $ \text{Tr} (\rho^{T_B} X ) =  \text{Tr} (\rho X^{T_B} ) $, we have 
\begin{align}
[\text{PT} (V)+ \frac{i}{2} \text{PT} (\Omega) ]_{jk} &= \langle (\xi_j\xi_k)^{T_B} \rangle  - \langle \xi_j^{T_B} \rangle \langle \xi_k^{T_B} \rangle ,\label{pttransformed2}
\end{align}
For the quantum optical case the PT operation on the operators gives the transformation $ \xi^{T_B} = (x_A, p_A, x_B, -p_B) $ \cite{braunstein}.  We note that for two operators both on $ B$ the transpose requires interchange of the order $ (X_B Y_B)^{T_B} = Y_B^{T_B} X_B^{T_B} $.  Simon's well-known criteria in terms of the submatrices of $ V $ (Eq. (17) of Ref. \cite{simon}) is equivalent then to finding the eigenvalues of the left hand side of (\ref{necessary}), and checking for any negative eigenvalues.  

This generalizes the covariance matrix formalism to an arbitrary set of operators.  We summarize the procedure here for convenience: (I) Choose a set of observables (\ref{xivector}) and calculate the partial transpose operators  $ (\xi_j\xi_k)^{T_B} $ and $  \xi_j^{T_B} $.  (II) Perform measurements such that the $N \times N $ matrix (\ref{pttransformed2}) can be constructed.  (III) Evaluate eigenvalues of (\ref{pttransformed2}), any negative value indicates entanglement.  
A violation of (\ref{necessary}) is only a sufficient condition for entanglement, hence it is possible that an entangled system can yield positive eigenvalues.  Thus the choice of operators is in this sense important. Specifically, it is important to obtain a non-zero $ \text{PT} (\Omega) $ as the positivity of $ \text{PT} (V) $ alone is guaranteed.  On the other hand there is a great flexibility in the procedure as all the available measurements can be put into the covariance matrix, and it is not necessary to put the variables in a particular basis that is suitable for detecting entanglement.

\section{Examples}
\begin{figure}[t]
\includegraphics[width=\linewidth]{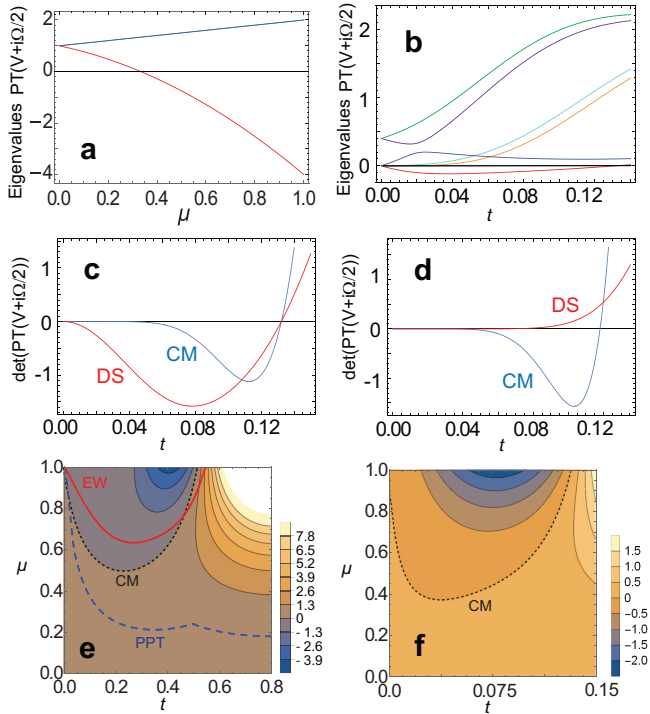} 
\caption{Covariance matrix detection of entanglement for spin systems. (a) Eigenvalues of the matrix (\ref{pttransformed2}) for the Werner-Bell state (\ref{mixed}) with operators $ \xi = \{ \sigma_A^x \sigma_B^x, \sigma_A^y \sigma_B^y, \sigma_A^z \sigma_B^z \} $. (b) Eigenvalues, (c) determinant of the  matrix (\ref{pttransformed2}) for the state (\ref{szszstate}) with $ \mu =1  $  and $ M = 20 $ with operators (\ref{xispin}). In (c), the covariance matrix (CM) result is compared to the Duan-Simon (DS) approach \cite{Duan,simon}.  (d)  Same as (c) but with using spin variables $ {S^x}' = (S^x+S^y)/\sqrt{2},  {S^y}' = (S^y-S^x)/\sqrt{2}, {S^z}' = S^z $. (e) The determinant of (\ref{pttransformed2}) for the mixed state (\ref{mixed}) for the state (\ref{szszstate}) with $ M = 2 $.  The boundary for the PPT criterion and the entanglement witness (EW) is shown for comparison, where entanglement is detected above the line.  (f) Same as (e) but for $ M = 20 $.}
\label{fig} 
\end{figure}

We illustrate our covariance matrix based entanglement criteria by looking at some 
examples. Towards this end we consider states of Werner form 
\begin{align}
\hat{\rho} = \frac{1-\mu}{D}I+\mu|\Psi\rangle\langle\Psi|,
\label{mixed}
\end{align}
where $\mu$ is the mixing parameter and $|\Psi \rangle$ is a pure state. For the first example we consider two qubits in a Bell state 
$ |\Psi\rangle = (|0\rangle | 0 \rangle + |1 \rangle | 1 \rangle )/\sqrt{2} $ with $D=4$,
for which the $N=3$ operators are $ \xi = \{ \sigma_A^x \sigma_B^x, \sigma_A^y \sigma_B^y, \sigma_A^z \sigma_B^z \} $.  
A PT operation changes the sign of only $ (\sigma_B^y)^{T_B} = -\sigma_B^y  $ and leaves all the other Pauli operators invariant. Since the product of two Pauli operators give either the identity or another Pauli operator, the only correlations to be measured are $ \langle \xi_j \rangle $ up to a sign.  The resulting eigenvalue spectrum of (\ref{pttransformed2}) is shown in  Fig. \ref{fig}(a).  One negative eigenvalue is present for $1/3 \le \mu \le 1$, detecting the full range of entangled Werner-Bell states \cite{nielsen2001separable}.  

As mentioned in the introduction the main motivation of our work is to detect entanglement in high dimensional
systems using only a limited number of correlations.  We thus next consider a Werner state (\ref{mixed}) of two entangled spin ensembles $A$ and $B$
\begin{align}
|\Psi\rangle = e^{i S^{z}_A S^{z}_B t} | S^x_A = M \rangle  | S^x_B = M \rangle 
\label{szszstate}
\end{align}
where $ S^{x,y,z}=\sum_{l=1}^M \sigma_l^{x,y,z} $, 
with $M$ being the number of qubits in the ensemble (taken to be the same for both).  
Here, $ | S^x_{A,B} = M \rangle $ are maximally polarized spin states in the $ S^x $ direction and the symmetric Hilbert space dimension is $ D = (M+1)^2 $.  Entanglement in one \cite{busch2014,sorensen,korbiczspin} and two spin ensembles \cite{byrnes2013,kurkjian2013spin,burlak2009entanglement,Byrnes2015102,ilookeke2014,pyrkov14,
giovannetti2003characterizing} has been well-studied in the context of quantum metrology and information.  
The pure state (\ref{szszstate}) is entangled at all times except for the disentangling times at $ t = \pi n /2 $ 
for integer $ n $ \cite{byrnes2013,kurkjian2013spin}. Although several entanglement criteria have 
been introduced for such spin systems \cite{busch2014,sorensen,korbiczspin,byrnes2013,kurkjian2013spin}, up to now no general procedure for 
entanglement detection using covariance matrices has been developed.

We consider the scenario where only second order correlations 
of the total spin operators can be measured.  The $N=6$ operators we consider are 
\begin{align}
\xi=(S^{x}_A,S^{y}_A,S^{z}_A,S^{x}_B,S^{y}_B,S^{z}_B).
\label{xispin}
\end{align}
Similar to the qubit example considered above we find that only  $ (S^{y}_B)^{T_B} = -S^{y}_B  $ undergoes a sign 
change under a partial transpose, and other spin operators are invariant.  The eigenvalue spectrum of the pure state (\ref{szszstate}) is given in Fig \ref{fig}(b) for $M = 20$. 
From our results we observe that there is one negative eigenvalue in the time range 
$ 0 < t \lesssim 0.13 $, thus in this case entanglement is detected only for a limited time range.  This is expected as only second order correlations are used, not the full information of the density matrix.  From the observation that there is only one negative eigenvalue, an equivalent method of detecting entanglement is to evaluate the 
determinant (the product of all eigenvalues) of the matrix (\ref{pttransformed2}), and test for positivity.  The determinant approach is 
shown in Fig \ref{fig}(c), where we observe that the same region of entanglement as shown in 
Fig \ref{fig}(b) is detected.  

We compare the performance of our method by performing a Holstein-Primakoff approximation where the spin operators are treated as approximate position and momentum operators $ x_{A,B} \approx S^y_{A,B} /\sqrt{2M}, p_{A,B} \approx S^z_{A,B}/\sqrt{2M} $.  This can then be applied to continuous variable methods such as the criterion of Duan, Simon, and co-workers \cite{Duan,simon}.  From the results in 
Fig \ref{fig}(c), we find that the Duan-Simon criteria detects the same region of entanglement as our 
covariance method.  This can be explained by the fact that when the operators are chosen to be field quadratures, our formulation reduces to Simon's criterion \citep{simon}, and thus are equivalent. While we have a $ 6 \times 6 $ matrix rather than a $ 4 \times 4 $ in Simon's criterion, the position and momentum 
operators $ x_{A,B},p_{A,B} $ are already optimized for the correlations that are produced by the state, and the additional two operators do not add further information. On the other hand, the measured operators may be quite different from the optimal operators in an experiment, due to the 
incomplete knowledge of the state, or potential calibration errors.  A quantum state is characterized in a 
typical experiment by measuring the various correlations between observables.  In such cases another 
choice of operators $ x_{A,B}',p_{A,B}' $ 
will then sub-optimally  detect entanglement using the Duan-Simon approach as can be seen from Fig. \ref{fig}(d).  
In contrast our criterion still continues to detect entanglement since no information of the optimal 
basis needs to be input to the covariance matrix.  This comes about due to the symmetric 
set of operators (\ref{xispin}) which span the same operator space under SO(3) rotations.  

Turning to mixed states, a comparison between PPT criteria, our covariance matrix method and the entanglement witness method \cite{horedecki1996separability,PhysRevA.62.052310,szangolies2015detecting,singh2016entanglement} for 
$M=2$ is shown in Fig. \ref{fig}(e).  The entanglement witness approach numerically optimizes a witness operator $ W = \sum_{ij} c_{ij} a_i \otimes b_i $ over coefficients $c_{ij} $, where $ a_i, b_i \in \{ I, S^x_{A,B}, S^y_{A,B}, S^z_{A,B} \} $  \footnotemark[53].  We see that the covariance matrix method and the entanglement witness 
method detect a smaller region of entanglement in contrast to the PPT criterion.  This is expected again because 
the PPT criterion uses the full density matrix, but the other two methods use only a limited set of 
correlators.  We find that the covariance matrix 
method can detect entanglement in a larger region of the state space than the entanglement witness approach, and in particular is more efficient in detecting entanglement when the state is more mixed (i.e. smaller $\mu$).  Further we notice that even for this small Hilbert space dimension $D=9$, computing the 
entanglement witness is numerically intensive, due to the constrained optimization involved in the 
procedure.  This is in contrast to our method which requires evaluating a determinant of 
(\ref{pttransformed2}) which can be done in a time polynomially scaling with $N$.

For larger systems $ M = 20 $, the covariance matrix method continues to detect entanglement for times in the region $ 0 < t \lesssim 1/2\sqrt{M} $ \cite{jing18}.  A comparison to the PPT criterion shows that to the accuracy of the plot the full range $ 0 \lesssim \mu \le 1 $, $ 0 < t \le 0.15 $ exhibits entanglement.  The reason for the large region of entanglement is that for systems in a larger Hilbert space, the amount of entanglement is accordingly larger, and are of a form that is relatively robust \cite{byrnes2013}. Again the limited region of entanglement detection is the price to be paid of only the low order correlations.  Despite the increased complexity of the state,  our method continues to detects a large portion of the parameter space.  We find that in this case the dimension of the Hilbert space already makes the 
entanglement witness method unreliable due to incomplete optimization.  This is again due to the complexity of the optimization which scales with the Hilbert space dimension $ D $, rather than the number of operators $N$ chosen.

\section{Conclusions}
In summary, we have generalized the covariance matrix approach for detecting entanglement in bipartite systems.  Our approach allows for an arbitrary number and type of operators to be used to form a covariance matrix.    The symmetric formulation of the criteria (\ref{necessary}) in terms of matrix invariances automatically takes into account of symmetries that are present in the problem.  This means that it is not necessary to construct suitable observables that are sensitive to the particular entangled state that is being detected, and one can work in a basis-independent way. This suggests that the larger the number of 
observables that are included in the covariance matrix, better the chance of detecting the entanglement
as it includes further correlations that may be relevant to entanglement. We expect that the most promising application of the approach will be to high dimensional systems where complete tomography of the state is impossible or impractical.  In this case, one would like to detect entanglement using only partial information of the state, with a limited set of observables.  Our method is highly efficient in the sense that the complexity of calculating the entanglement is related to the number of operators chosen $N $, rather than the dimension of the Hilbert space $ D $.   We expect that the system can be applied to many different systems beyond those explored here, such as in quantum many-body systems where only a limited number of correlation functions are available.

\section{acknowledgments}
We thank Bartosz Regula for motivating this work, and Jonathan Dowling, Yumang Jing for discussions.  This work is supported by the Shanghai Research Challenge Fund; New York University Global Seed Grants for Collaborative Research; National Natural Science Foundation of China (61571301); the Thousand Talents Program for Distinguished Young Scholars (D1210036A); and the NSFC Research Fund for International Young Scientists (11650110425); NYU-ECNU Institute of Physics at NYU Shanghai; the Science and Technology Commission of Shanghai Municipality (17ZR1443600); and the China Science and Technology Exchange Center (NGA-16-001).

\section{APPENDIX}
\subsection{Symmetries of the matrices $ V $ and $ \Omega $ }

In this section we show the symmetry properties of the matrices $ V $ and $ \Omega $.  We first show that $ V $ is a real symmetric matrix:  
\begin{align}
V_{jk}^* & = \frac{1}{2} \langle  \{ \Delta \xi_j, \Delta \xi_k \}^\dagger \rangle \nonumber \\
& = \frac{1}{2} \langle \left( \Delta \Delta \xi_k^\dagger \Delta \xi_j^\dagger +  \xi_j^\dagger \Delta \xi_k^\dagger \right) \rangle  \nonumber \\
& = \frac{1}{2} \langle \{ \Delta \xi_j, \Delta \xi_k \} \rangle \nonumber \\
& = V_{jk} = V_{kj} ,
\end{align}
where we have used the fact that the chosen operators are Hermitian $ \xi_n^\dagger = \xi_n $.  

Next, we show that $ \Omega $ is a real antisymmetric matrix:
\begin{align}
\Omega_{jk}^* & = i \langle [  \xi_j, \xi_k ]^\dagger  \rangle \nonumber \\
& = i \langle  \left(  \xi_k^\dagger \xi_j^\dagger - \xi_j^\dagger \xi_k^\dagger \right) \rangle \nonumber \\
& = i \langle [ \xi_k, \xi_j ] \rangle  \nonumber \\
& = - \Omega_{kj} = \Omega_{jk} ,
\end{align}
which proves the statement.  We note that the matrices are defined in the $ N \times N $ space of the chosen operators $ \xi_n $, and not in the Hilbert space of the operators. Thus while the quantity $ -i [ \xi_j, \xi_k ] $ is Hermitian,  the matrix $ \Omega_{jk} $ is not.  

It then follows that the quantity in (4) of the main text is a Hermitian matrix:
\begin{align}
\left(V + \frac{i}{2} \Omega \right)^\dagger = V^T - \frac{i}{2} \Omega^T = V + \frac{i}{2} \Omega.
\end{align}
One can view $ V$ as being the real part and $ \Omega/2 $ as being the imaginary parts of the Hermitian matrix.

\subsection{Validity of (4) for mixed states}

In this section we show that (4) in the main text holds for mixed states.  Taking a general mixed state
\begin{align}
 \rho = \sum_l p_l | \Psi_l \rangle \langle  \Psi_l |, 
\end{align}
and using the fact that
\begin{align}
V_{jk}+\frac{i}{2}\Omega_{jk} = \langle \xi_j \xi_k \rangle - \langle \xi_j \rangle \langle \xi_k \rangle 
\end{align}
we have
\begin{align}
V_{jk} + \frac{i}{2} \Omega_{jk}  = & \sum_l p_l \langle \Psi_l |  \xi_j \xi_k |  \Psi_l \rangle  \nonumber \\
& - \sum_l \sum_{l'} p_l p_{l'} \langle \Psi_l |  \xi_j  |  \Psi_l \rangle \langle \Psi_{l'} |   \xi_k |  \Psi_{l'} \rangle . 
\end{align}
Given a positive semi-definite matrix, all the principal submatrices are also positive semi-definite. Even though our operators  $\xi_i$ do not form the complete basis in Hilbert space, they do correspond to one of the principal submatrices of the matrix formed by complete set of operators.  This enables us to use Eq. (29) of Gittsovich et.al.\cite{gitts} and the second term can be bounded as a matrix inequality in the labels $ j,k $ as 

\begin{align}
V_{jk} + \frac{i}{2} \Omega_{jk} & \ge \sum_l p_l \left( \langle \Psi_l |  \xi_j \xi_k |  \Psi_l \rangle  -\langle \Psi_l |  \xi_j  |  \Psi_l \rangle \langle \Psi_l |   \xi_k |  \Psi_l \rangle   \right). 
\end{align}
The quantity in the brackets is the pure state result as shown in (8) of the main text.  Since this is a positive semi-definite matrix, the probabilistic sum can be bounded by zero, which yields (4) for mixed states.

\subsection{Generalized uncertainty relations}

We start with defining a phase space vector $\xi=(\xi_1,\xi_2,..\xi_N)$ such that 

\begin{align}
[\xi_a,\xi_b]=i\Omega_{ab},
\label{omega}
\end{align}
where $a,b=1,2,..N$.  Here $\Omega_{ab}$ is the $N$ ${\times}$ $N$ dimensional commutation matrix. We give a general formalism for $N$ arbitrary operators $\{\hat{\xi}_i\}$. Defining  
\begin{align}
|f_i\rangle=(\hat{\xi_i}-\langle\hat{\xi_i}\rangle)|\Psi\rangle  ,
\end{align}
the variance in various operators is 
\begin{align}
\sigma^2_i & =\langle\Psi|(\hat{\xi_i}-\langle{\hat{\xi_i}}\rangle)^2|\Psi \rangle =\langle{f_i}|f_i\rangle .
\end{align}
The product of all such operators is then given as
\begin{align}
\prod_{i}\sigma^2_i & =\prod_{i}\langle{f_i}|f_i\rangle.\label{product}
\end{align}
Given a set of $N$ state vectors $\{|f_{i}\rangle;i=1,2,..,N\}$, we can use the Gram-Schmidt procedure \cite{arfken} to obtain a set of $N$ orthogonal (but unnormalized) state vectors as
\begin{align}
|f^{\perp}_i\rangle=|f_i\rangle-\sum_{j=1}^{i-1}\frac{\langle{f}^{\perp}_{j}|f_i\rangle}{\langle{f}^{\perp}_{j}|f^{\perp}_j\rangle}|f^{\perp}_j\rangle
\end{align}
Now using the positivity of norm of the orthogonal states, (\ref{product}) gives for $N=1$
\begin{align}
 & \sigma^2_{\xi_1}  \geqslant 0  \label{1} .
 \end{align}
 For $N=2$, we use the Cauchy Schwarz inequality which is defined as
 \begin{align}
 &  \langle{f_1}|f_1\rangle\langle{f_2}|f_2\rangle \geqslant |\langle{f_1}|f_2\rangle|^2  \label{1a}
 \end{align}
to obtain Schrodinger's uncertainty relation, 
\begin{align}
\sigma^2_{\xi_1}\sigma^2_{\xi_2} \geqslant & \abs{\frac{1}{2}\langle\{\hat{\xi_1},\hat{\xi_2}\}\rangle-\langle \hat{\xi_1}\rangle\langle \hat{\xi_2}\rangle}^2+\abs{\frac{1}{2i}\langle[\hat{\xi_1},\hat{\xi_2}]\rangle}^2 
\label{2}
\end{align} 
For $N=3$, we obtain the three operator uncertainty relation which in terms of the state vectors can be given as 
\begin{align}
\sigma^2_{\xi_1}\sigma^2_{\xi_2}\sigma^2_{\xi_3}   \geqslant &    \langle{f_1}|f_1\rangle|\langle{f_2}|f_3\rangle|^2 + \langle{f_1}|f_2\rangle\langle{f_2}|f_3\rangle\langle{f_3}|f_1\rangle          & \nonumber \\ 
&  +\langle{f_2}|f_2\rangle|\langle{f_3}|f_1\rangle|^2 +\langle{f_2}|f_1\rangle\langle{f_3}|f_2\rangle\langle{f_1}|f_3\rangle   &\nonumber \\
&+\langle{f_3}|f_3\rangle|\langle{f_1}|f_2\rangle|^2 &  
  \label{ee}
\end{align}
which gives the expression in the main text. \\\\

\subsection{Entanglement witness approach}

In this section we detail how to calculate the boundary as shown in Fig. 1(e) using an entanglement witness \cite{horedecki1996separability,PhysRevA.62.052310} approach.  The method we follow is an optimization approach as described in Ref. \cite{szangolies2015detecting}. We assume the situation where only the correlations $ \langle a_i \otimes b_j \rangle $
are available, where the operators on the $ A $ and $ B $ subsystems are 
\begin{align}
a_i & \in \{ I_A, S_A^x,  S_A^y,  S_A^z) \nonumber \\
b_i & \in \{ I_B, S_B^x,  S_B^y,  S_B^z)
\end{align}
respectively.  We then define the quantity
\begin{align}
W = \sum_{ij=0}^3 c_{ij} a_i \otimes b_i
\label{woperator}
\end{align}
where $ c_{ij} $ are real coefficients.  The procedure is then 
\begin{align}
\text{Minimize } \langle W \rangle \hspace{3mm} & \nonumber \\
 \text{Subject to: (1) } & W = P + Q^{T_A} \nonumber \\
 \text{(2) } & P \ge 0  \nonumber \\
 \text{(3) } &Q \ge 0 \nonumber \\
 \text{(4) } & \text{Tr}(W) = 1 .
\end{align}
The only term with non-zero trace in (\ref{woperator}) is the coefficient of $ I_A \otimes I_B $, and the normalization condition is satisfied if
\begin{align}
c_{00} = \frac{1}{(N+1)^2} .
\end{align}
The remaining coefficients $ c_{ij} $ are found by a simulated annealing random search procedure.

% How to do the references:
%% 1) First uncomment the below and compile
%\bibliographystyle{apsrev}
%\bibliographystyle{unsrt}
%\bibliography{NIIReferences}
%% 2) Copy the .bbl file to below and comment out the above two lines.

\end{document}